%% file: main.tex
\documentclass[9pt,sigconf]{acmart}
\usepackage[shortcuts]{extdash} 
\usepackage{booktabs} %

\graphicspath{{figure/}{figures/}}

\usepackage{ifthen}
\newboolean{showcomments}
\setboolean{showcomments}{true}
\ifthenelse{\boolean{showcomments}}
{ \newcommand{\mynote}[3]{
    \protect\fbox{\bfseries\sffamily\scriptsize#1}
    {\small$\blacktriangleright$\textsf{\emph{\color{#3}{#2}}}$\blacktriangleleft$}}}
{ \newcommand{\mynote}[3]{}}

\newcommand{\dku}[1]{\mynote{Dirk}{#1}{blue}}

\setcopyright{none} 

\begin{document}
\title{Statement: The Metaverse as an \\Information-Centric Network}
\titlenote{This paper has been accepted for publication in the proceedings of ACM ICN 2023. Please the cite the published version DOI:10.1145/3623565.3623761}

\author{Dirk Kutscher}
\orcid{0000-0002-9021-9916}
\affiliation{%
  \institution{HKUST (GZ)}
  \country{China} 
}
\email{dku@ust.hk}

\author{Jeff Burke}
\orcid{0000-0003-2240-5642}
\affiliation{%
  \institution{UCLA REMAP}
  \country{USA} 
}
\email{jburke@remap.ucla.edu}

\author{Giuseppe Fioccola}
\orcid{0000-0003-1885-5106}
\affiliation{%
  \institution{Huawei}
  \country{Italy} 
}
\email{giuseppe.fioccola@huawei.com}

\author{Paulo Mendes}
\orcid{0000-0003-1059-8272}
\affiliation{%
  \institution{Airbus}
  \country{Germany} 
}
\email{paulo.mendes@airbus.com}

\renewcommand{\shortauthors}{D. Kutscher, J. Burke, G. Fioccola, P. Mendes }

\begin{abstract}
This paper discusses challenges and opportunities of considering the Metaverse as an Information-Centric Network (ICN). 
The Web today essentially represents a data-centric application layer: data named by URLs is manipulated with REST primitives. However, the semantic gap with the underlying host-oriented transport is significant, typically leading to complexity, centralization, and brittleness. Popular interest in ``the Metaverse'' 
suggests that the end-user experience of the Web will evolve towards always-on \textit{eXtended Reality} (XR). 
With the benefit of a historical perspective, computing advances, and decades of experience with a global network, there is an opportunity to holistically consider the Metaverse not as an application of the current network, but an evolution of the network itself, reducing rather than widening the gap between network architecture and application semantics. 
An ICN architecture offers the possibility to achieve this with less
overhead, low latency, better security, and more disruption tolerance
suitable to diverse uses cases, even those facing intermittent
connectivity.
\end{abstract}



\maketitle

\input{intro}

\input{concept}
\input{req}

\bibliographystyle{acm}
\bibliography{main.bib}

\end{document}

%% file: intro.tex
\section{Introduction}
\label{sec:intro}

The Web today has a specific technical definition: it includes
presentation layer technologies, protocols, agreed-upon ways of
achieving certain semantics such as \textit{Representational State
  Transfer} (REST) \cite{fielding00}, and security
infrastructure. However, from a user perspective, it can be viewed as
a universe of consistently navigable content and (occasionally)
interoperable services. The user experience and architectural
underpinnings have evolved in parallel and have influenced each other:
for many end users, the Web and the network are synonymous. Rather
than building up ``Metaverse'' as an application domain based on IP,
we aim to explore ``the Metaverse'' as strongly intertwined with ICN,
just as the \emph{modern concept} of the Web and its \emph{technology
stack} are inseparable for a broad set of applications.

As a placeholder name for a range of new technologies and experiences,
``the Metaverse`` is even less well-defined than the Web. We adopt the
commonly used concept of a shared, interoperable \cite{Erickson_2023},
and persistent XR. Some descriptions and early prototypes for social
AR/VR systems \cite{10.1145/3517745.3561417} suggest leveraging
existing Internet and Web protocols to provide Metaverse services,
without addressing the technical complexity and centralization of
control required to provide the underlying cloud service
infrastructure \cite{Lamina1}.
Here, we do not take as given current designs and deployment models
that consider the Metaverse as an overlay application with
corresponding infrastructure dependencies, as this exacerbates the
current gaps (and the resulting costs and technical complexity)
between distributed applications and the underlying network
architecture.  Instead, we assume a fundamentally
information\-/centric system in which most applications participate in
granular 3D content exchange, context-aware integration with the
physical world, and other Metaverse-relevant services.
``The Metaverse'' \emph{is} an information-centric concept that likely will
become synonymous with the network itself.
We argue that reciprocal design of the network and applications
will open new opportunities for the deployment of
Metaverse-suggestive experiences even today.

%
%

%% file: concept.tex
\section{Concept}
\label{sec:concept}

Experientially, this Metaverse is an extension of the Web into
immersive XR modalities that are often aligned with physical space, as in augmented reality (AR).
We conceive the Metaverse not only as a shared XR environment, but the
next generation of the web, extending into 3D interaction/immersion
and optionally overlaid on physical spaces.  Instead of rendering data
objects into a 2D page (within a tab within a window) on a device, we
envision such objects being rendered into a shared 3D space,
interacting among each other and with end users.
Architecturally, leveraging ICN concepts provides support for
decentralized publishing, content interoperability and co-existence,
based on general building blocks and not within separated application
silos as today's initial prototypes. We claim that such properties are
required to achieve the generally circulated visions of Metaverse
systems, but are not achievable today because of the host- and
connection-centric way in which the web operates and is presented to
users in browsers.  We point out four ICN capabilities critical to
Metaverse concepts:
i) scalable and robust multi-destination communication, overcoming IP
multicast challenges \cite{moore2023ruined}, such as inter-domain
routing, scalability, and routing communication overhead; ii)
leveraging wireless broadcast to support shared local views and
low-latency interactivity without application-awareness in edge
routers; iii) privacy, selective attention, content filtering, and
autonomous interactions, as well as ownership and control on the
publishing side; and iv) supporting in-network processing for objects
replication and transformation.

For example, imagine {\em interactive holographic communication}
consisting of participants' 3D video, spatial audio, and shared 3D
documents.  In ICN, such an application can represent virtual content
as secure data objects and share them efficiently in a larger group of
peers, fetching only the data necessary to reconstruct a suitable
representation while being aware of the constraints of user devices
and access networks. Furthermore, while experiencing 3D objects shared
by the group, each participant may also interact in the same XR
environment with personal services such as wayfinding, messaging, and
\emph{Internet of Things} (IoT) device status.  Interactions between
private and shared 3D objects
would be simplified if these objects use similar conventions but with
different security. This concept is semantically well-aligned with ICN
properties, particularly for security, as it revolves around
object-level data exchange rather than hosts or channels.
Integration and interoperability within a shared XR environment,
without centralization, is challenging if one has to negotiate not
only data interactions but also the underlying service connections and
security relationship using host-centric paradigms.  It also
exacerbates the impact of intermittent connectivity on interactivity
when the global network is required for functions such as rendezvous
-- that are handled locally in ICN.

As a second example, consider {\em creating a shared environment} --
e.g., to pre-visualize engineering models of an aircraft -- from a
collection of collaboratively edited 3D documents.  Imagine
component documents interacting in a simulation.
Documents can be modularized, linked, and overlaid in a web-like
manner. Today, such cross-platform interoperability and visualization
without centralized hubs is impractical \cite{Radoff_2022}, and
it is difficult to create secure, granular data flows
required for interaction between co-existing 3D elements to ``bring
them to life'' in a virtual world. In an ICN approach, such modules
could be independently authored and published, shared between
applications, becoming building blocks of a richer,
\emph{interacting} system of user- and machine-generated content.

%% file: req.tex
\section{Technical Challenges}
\label{sec:req}

Many applications already employ data-oriented paradigms.
Mapping them to a host-centric network model creates complexities and
robustness issues that can be addressed with using a native ICN
approach. While ICN benefits and deployment challenges have been
discussed extensively in the literature, we focus on the unique
research challenges and opportunities for ICN to be the underlying
fabric of a web-like Metaverse.
Two key examples: i) Approaches to ``interconnect'' virtual objects
and systems published by different owners.  For instance, implementing
pouring a virtual cup of tea from a kettle owned by one user (and service) into another user's cup (hosted by another service)
\cite{patil22:kua}; and ii) ICN versions of emerging XR object and
communication standards, e.g., \textit{Universal Scene Description}
(USD)~\cite{usd} and glTF~\cite{gltf}: ICN can be used for consistent
and efficient sharing of scene and model descriptions. In addition,
these and other key application-layer XR data structures are based on
object hierarchies. Such documents can be hyper-media objects linking
multiple components into a large context. ICN can make this a natural
approach, operating on a fine-grained basis.  Research is also
required on communication and security paradigms, in particular for
mutable versions of these objects.

Low-latency exchange of arbitrary objects and data streams is a
fundamental enabler for Metaverse applications. This represents a
challenge for current CDN-based infrastructures based on HTTP services
such as DASH-based video on-demand streaming. For interactive
multimedia, WebRTC protocols are more suitable, but they shift
significant complexity to applications and do not provide a
cross-application way to exchange data objects. A transition from the
abstraction of ``streaming'' to selectively shared state may be much
more suitable for Metaverse applications and is potentially well
supported by distributed dataset synchronization techniques in
ICN~\cite{moll2022sok}.
Current overlay approaches, such Media over QUIC (MoQ)
\cite{I-D.ietf-moq-requirements} and extensions such as QuicR
\cite{I-D.jennings-moq-quicr-proto}, blend real-time interactive media
with streaming, albeit with some complexity.

What is needed is a fine-grained, hierarchical media exchange for
low-latency interactive communication that enables scalable
multi-destination distribution and in-network replication and
transformation that {\bf exposes application object hierarchy for fine-grained retrieval
and security.} To support denser, large-scale communication sessions,
such a service should be able to seamlessly leverage wireless
broadcast.  It must also provide support for heterogeneous devices and
edge networks, i.e., providing only data elements needed for
rendering, at different quality layers, possibly leveraging dynamic
transcoding and level-of-detail support. With respect to low-latency
and QoS in ICN, we suggest further research and experiments on {\bf
  fine-tuning interest aggregation, caching and its influence on
  receiver-based performance estimation}, and the development of {\bf
  specific QoS mechanisms} \cite{rfc9064} to prioritize critical
requests, such as prioritizing interaction data and
baseline-quality media objects over higher quality objects.
The fine-granular distribution, on-demand creation, and sharing of
data objects in such systems calls for suitable security solutions
that go beyond ICN's current object signing and encryption
mechanisms. NDN Trust Schemas \cite{10.1145/2810156.2810170}
demonstrated automated fine-granular authorization. In addition,
mechanisms such as {\bf provenance verification for data
  transformation} and UI support for {\bf visualizing and
  authenticity and provenance} are needed as well.

In conclusion, we encourage consideration of the close relationship
between the end-user experience of a Metaverse and an ICN
architecture.
An information-centric Metaverse could not only enable interoperable
communication and data sharing, but also become the lingua franca for
internal object representation and composition in platforms such as 3D
game engines, similar to current Internet building blocks such as
TCP, HTTP, and DNS that are used within application frameworks and
microservice platforms.